\documentclass[12pt,letter]{article}

\usepackage{graphicx, epsfig, color}
\def\to{\rightarrow}

\def\bi{\begin{itemize}}
\def\ei{\end{itemize}}

\def\sps1ap{SPS1a$^\prime$}
\def\c1p{C1$^\prime$}

\def\tb{\tilde b}

\def\tst{\tilde t}

\def\tg{\tilde g}

\def\tq{\tilde q}
\def\tw{\widetilde W}
\def\tz{\widetilde Z}
\def\alt{\stackrel{<}{\sim}}
\def\agt{\stackrel{>}{\sim}}
\def\be{\begin{equation}}  
\def\ee{\end{equation}}  
\def\bea{\begin{eqnarray}}  
\def\eea{\end{eqnarray}}  
\def\beas{\begin{eqnarray*}}  
\def\eeas{\end{eqnarray*}}  
\newcommand\prd[3]{{\it Phys.\ Rev.\ }{\bf D #1} (#2) #3}

\newcommand\prl[3]{{\it Phys.\ Rev.\ Lett.\ }{\bf #1} (#2) #3}
\newcommand\plb[3]{{\it Phys.\ Lett.\ }{\bf B #1} (#2) #3}
\newcommand\jhep[3]{{\it J. High Energy Phys.\ }{\bf #1} (#2) #3}

\newcommand\npb[3]{{\it Nucl.\ Phys.\ }{\bf B #1} (#2) #3}
\newcommand\epjc[3]{{\it Eur.\ Phys.\ J. }{\bf C #1} (#2) #3}
\newcommand\ptp[3]{{\it Prog.\ Theor.\ Phys.\ }{\bf #1} (#2) #3}

\newcommand{\hepph}[1]{hep-ph/#1}

\newcommand\ppnp[3]{{\it Prog.\ Part.\ Nucl.\ Phys.}{\bf  #1} (#2) #3}

\begin{document}
\begin{titlepage}

\vspace{0.5cm}
\begin{center}
{\Large \bf SUSY models under siege:\\
LHC constraints and electroweak fine-tuning
}\\ 
\vspace{1.2cm} \renewcommand{\thefootnote}{\fnsymbol{footnote}}
{\large Howard Baer$^1$\footnote[1]{Email: baer@nhn.ou.edu },
Vernon Barger$^2$\footnote[2]{Email: barger@pheno.wisc.edu },
Dan Mickelson$^1$\footnote[3]{Email: dmickelso@nhn.ou.edu},\\
and Maren Padeffke-Kirkland$^1$\footnote[4]{Email: m.padeffke@ou.edu } 
}\\ 
\vspace{1.2cm} \renewcommand{\thefootnote}{\arabic{footnote}}
{\it 
$^1$Dept. of Physics and Astronomy,
University of Oklahoma, Norman, OK 73019, USA \\
}
{\it 
$^2$Dept. of Physics,
University of Wisconsin, Madison, WI 53706, USA \\
}

\end{center}

\vspace{0.5cm}
\begin{abstract}
\noindent 
Recent null results from LHC8 SUSY searches along with the discovery of a SM-like 
Higgs boson with mass $m_h\simeq 125.5$ GeV indicates sparticle masses in the TeV range, 
causing tension with conventional measures of electroweak fine-tuning.
We propose a simple {\it Fine-tuning Rule} which should be followed under any credible 
evaluation of fine-tuning. We believe that overestimates of electroweak fine-tuning 
by conventional measures all arise from violations of this rule.
We show that to gain accord with the Fine-tuning Rule, 
then both Higgs mass and the traditional $\Delta_{BG}$ 
fine-tuning measures reduce to the model-independent electroweak fine-tuning measure $\Delta_{EW}$.
This occurs by combining dependent contributions to $m_Z$ or $m_h$ into independent units.
Then, using $\Delta_{EW}$, we evaluate EW fine-tuning for a variety of SUSY models including 
mSUGRA, NUHM1, NUHM2, mGMSB, mAMSB, hyper-charged AMSB and nine cases of 
mixed moduli-anomaly (mirage) mediated
SUSY breaking models (MMAMSB) whilst respecting LHC Higgs mass and $B$-decay constraints 
(we do not impose LHC8 sparticle mass constraints due to the possibility of compressed spectra
within many of these models).
We find mSUGRA, mGMSB, mAMSB and MMAMSB models all to be highly fine-tuned. 
The NUHM1 model is moderately fine-tuned while NUHM2 which allows for radiatively-driven naturalness (RNS)
allows for fine-tuning at a meager 10\% level in the case where $m(higgsinos)\sim 100-200$ GeV and the 
TeV-scale top squarks are well-mixed. Models with RNS may or may not be detect at LHC14.
A $\sqrt{s}\sim 500$ GeV $e^+e^-$ collider will be required to make a definitive search for the 
requisite light higgsinos.

\vspace*{0.8cm}

\end{abstract}

\end{titlepage}

\section{Introduction}

It has long been claimed that electroweak naturalness requires that the superpartners of 
the SM fields exist with masses of order the weak scale\cite{eenz,bg,kane,ac1,dg,ccn,ellis2,king,casas,fp,Nomura:2005qg,ross,derm_kim,shafi,perel,antusch,hardy,sug19,Fichet:2012sn,Younkin:2012ui,Kowalska:2013ica,han,Dudas:2013pja,Arvanitaki:2013yja,Fan:2014txa,Gherghetta:2014xea,Kowalska:2014hza,feng,ltr,rns,comp,deg,Martin:2013aha,Fowlie:2014xha,azar_xt}
$m(sparticles)\sim m_{weak}\sim m_Z$.
Already at LEP2, the lack of signal for chargino pairs called into question whether there might exist
a ``Little Hierarchy Problem''\cite{barbstrum} characterized by $m(sparticle)\gg m_Z$.
This viewpoint has seemingly been strengthened by 
\bi
\item the lack of any signal for sparticles at LHC8\cite{atlas_susy,cms_susy} which requires $m_{\tg}\agt 1.8$ TeV for models
with $m_{\tq}\sim m_{\tg}$ and $m_{\tg}\agt 1.3$ TeV for models with $m_{\tq}\gg m_{\tg}$ and
\item the rather large value of $m_h\simeq 125.5$ GeV\cite{atlas_h,cms_h} 
which requires multi-TeV top squarks with small mixing or 
TeV-scale top squarks with large mixing\cite{mhiggs,hpr,h125,arbey}.
\ei
If indeed weak scale SUSY is highly fine-tuned in the electroweak sector, then likely SUSY 
is not as we know it since the twin requirements of parsimony and naturalness cannot be met 
simultaneously\cite{craig}.
Abandoning parsimony is not a step lightly taken since the further one strays from known physics the more
likely one is to be wrong.
But before jumping to such strong conclusions-- which may well guide support for and construction of future
HEP experimental facilities-- it is worthwhile to scrutinize the available measures of 
electroweak fine-tuning (EWFT) in SUSY models.
Indeed, in a recent paper we have claimed that {\it conventional measures tend to overestimate EWFT in 
supersymmetric models}, often by several orders of magnitude\cite{comp}.

In order to ascertain when a claim of fine-tuning is legitimate, we propose a simple
{\bf Fine-tuning Rule} which may act as a guide:
\begin{quotation}
{\it When evaluating fine-tuning, it is not permissible to claim fine-tuning of 
{\bf dependent} quantities one against another.}
\end{quotation}
We believe the over-estimates of EWFT by conventional measures referred to above all come from
violations of this rule.

To be explicit, most theories contain several, perhaps many, parameters. Some of these may be set 
equal to measured values, while others may be undetermined or at least constrained, but may vary over a
wide range of values. The parameters are frequently introduced to parametrize our ignorance of more fundamental 
physics, and their variation allows one to encompass a wide range of possibilities. 
We can think of each parameter as a dial, capable of being adjusted to specific, or alternatively a wide range of
values. If some contribution to a measured quantity ({\it e.g.} $m_h^2$ or $m_Z^2$ in this paper) in a theory blows up, 
and we have an adjustable parameter which may be dialed independently
to compensate, then we may legitimately evaluate fine-tuning: is a huge, unnatural cancellation required?
Alternatively, if as a consequence of one contribution blowing up, a related dial/parameter is driven to
large, opposite-sign compensating values, then any claimed fine-tuning would violate our rule (the
quantities would be {\it dependent}) and some regrouping of terms into independent quantities should be found. 
We will meet some clarifying examples in the subsequent sections of this paper.

\subsection{Simple electroweak fine-tuning}

In most supersymmetric models based on high scale input parameters-- {\it i.e.} 
SUSY models with soft term boundary conditions imposed at a scale $\Lambda\gg m_{weak}$ 
where $\Lambda$ may range as high as $m_{GUT}\simeq 2\times 10^{16}$ GeV or 
even the reduced Planck mass $M_P\simeq 2\times 10^{18}$ GeV--
the soft SUSY breaking terms are input at the scale $\Lambda$ and then evolved to the electroweak
scale $m_{weak}$ via renormalization group (RG) running.\footnote{Here we differentiate the
superpotential Higgs/higgsino mass term $\mu$ from the soft breaking terms, as do most model
builders, and we return to the SUSY $\mu$ problem later.} 
At the weak scale, the
scalar potential is minimized and checked to ensure that EW symmetry is properly broken. 
The value of $\mu$ is then fixed in terms of the weak scale soft SUSY breaking terms
$m_{H_u}^2$ and $m_{H_d}^2$ by requiring that the measured value of $m_Z\simeq 91.2$ GeV is obtained:
\be
\frac{m_Z^2}{2}=\frac{m_{H_d}^2 + \Sigma_d^d -
(m_{H_u}^2+\Sigma_u^u)\tan^2\beta}{\tan^2\beta -1} -\mu^2 \simeq -m_{H_u}^2-\Sigma_u^u-\mu^2
\label{eq:mZs}
\ee
where $\Sigma_u^u$ and $\Sigma_d^d$ are radiative corrections that
arise from the derivatives of $\Delta V$ evaluated at the minimum.
The radiative corrections $\Sigma_u^u$ and $\Sigma_d^d$ include contributions from
various particles and sparticles with sizeable Yukawa and/or gauge
couplings to the Higgs sector.
Expressions for the $\Sigma_u^u$ and $\Sigma_d^d$ are given in the Appendix of 
Ref.~\cite{rns}.

Already at this point: if $-m_{H_u}^2 (weak)$ in the right-hand-side of Eq. \ref{eq:mZs}
is large positive ($\gg m_Z^2$), then the value of $\mu$ must be fine-tuned by hand to ensure
the measured value of $m_Z^2$ is obtained. Since most researchers these days run automated computer
codes\cite{codes} to calculate the weak scale spectrum of SUSY and Higgs particles, this represents
a {\it hidden} fine-tuning that ought to be accounted for. 

\begin{figure}[tbp]
\includegraphics[height=0.4\textheight]{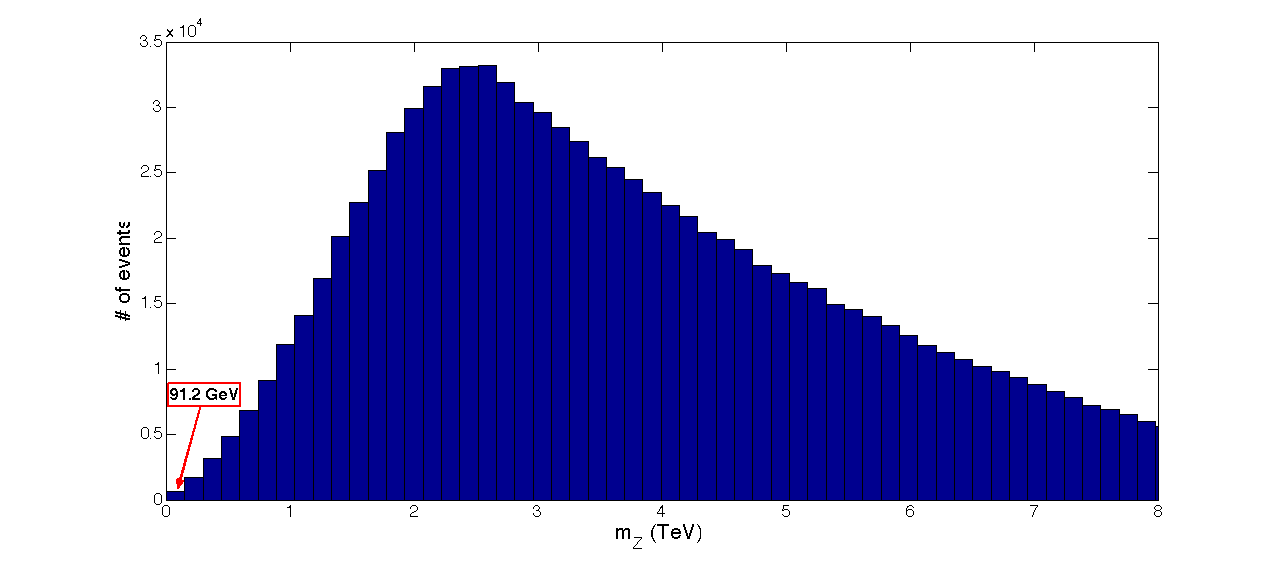}
\caption{Plot of value of $m_Z$ generated from a scan over
pMSSM model parameter space while {\it not} implementing the $m_Z^2$ constraint.
\label{fig:mz}}
\end{figure}

Alternatively, if soft SUSY breaking terms {\it and} $\mu$ are input parameters, then much 
higher values of $m_Z\gg 91.2$ GeV are expected from scans over SUSY model parameter space.
For example, in Fig. \ref{fig:mz} we plot the value of $m_Z$ which is generated from a 
scan over pMSSM parameter space\cite{pmssm}\footnote{The pMSSM, or phenomenological MSSM, is the MSSM
defined with weak scale input parameters where all CP violating and flavor violating soft terms have
been set to zero. Also, usually first/second generation soft terms are set equal
to each other to avoid flavor-violations.}. 
The 20 dimensional pMSSM parameter space then includes
\bea
M_1,\ M_2,\ M_3,\\
m_{Q_1},\ m_{U_1},\ m_{D_1},\ m_{L_1},\ m_{E_1},\\
m_{Q_3},\ m_{U_3},\ m_{D_3},\ m_{L_3},\ m_{E_3},\\
A_t,\ A_b,\ A_{\tau},\\
m_{H_u}^2,\ m_{H_d}^2,\ \mu,\ B.
\eea
The usual strategy is to use the EW minimization conditions\cite{wss} to trade the bilinear
parameter $B$ for the ratio of Higgs vevs $\tan\beta \equiv v_u/v_d$ and to exchange
$m_{H_u}^2$ and $m_{H_d}^2$ for $m_Z^2$ and $m_A^2$\cite{wss}. 
This procedure reduces the number of free parameters to 19 (since $m_Z$ is fixed) 
but hides the fine-tuning embedded in Eq. \ref{eq:mZs} since now $m_{H_u}^2$ is an output. 

Here we will avoid the $m_Z^2$ constraint and scan over the 20 dimensional pMSSM space for the range
of scalar and gaugino mass soft terms from $0-10$ TeV,
$-10\ {\rm TeV}< A_i<10$ TeV, $\mu :0-3$ TeV and $\tan\beta :3-60$,
while requiring the lightest neutralino $\tz_1$ as lightest SUSY particle (LSP) and $m_{\tw_1}>103.5$ GeV 
(in accord with LEP2 constraints).\footnote{This limit diminishes to
$\sim 91.9$ GeV in the case of a wino-like WIMP.}
Our results are shown in Fig. \ref{fig:mz}.
Here, we see that the most probable value of $m_Z$ is $\sim 2.5$ TeV with a large spread 
to both higher and lower values. It is highly unlikely to generate the measured value $m_Z=91.2$ GeV:
this is the essence of the Little Hierarchy problem.

\begin{figure}[tbp]
\includegraphics[height=0.4\textheight]{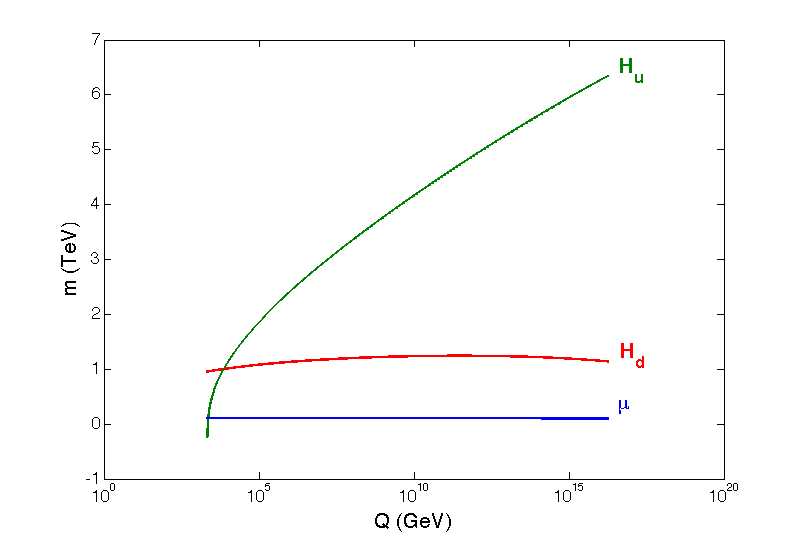}
\caption{Renormalization group evolution of 
$sign (m_{H_u}^2) \sqrt{|m_{H_u}^2|}$, $\sqrt{m_{H_d}^2}$ and $\mu$ 
versus energy scale $Q$ for the RNS benchmark point from Ref. \cite{ilcbm8}.
The value $m_A=1$ TeV $\simeq m_{H_d}(weak)$ and $\mu (weak) =110$ GeV.
\label{fig:Qrun}}
\end{figure}

Alternatively, the fact that $m_Z=91.2$ GeV along with $m_h\simeq 125.5$ GeV tells us from
Eq. \ref{eq:mZs} that to naturally generate the measured value of $m_Z$ (and $M_W$) and $m_h$, then
\bi
\item $|\mu|\sim m_Z\sim 100-200$ GeV
\item $m_{H_u}^2$ should be driven to small negative values such that $-m_{H_u}^2\sim 100-200$ GeV
at the weak scale and
\item that the radiative corrections are not too large: $\Sigma_u^u\alt 100-200$ GeV
\ei
The first two of these conditions are shown in Fig. \ref{fig:Qrun} as 
soft term and $\mu$ RG running versus $Q$ for a
radiatively-driven natural SUSY benchmark point from Ref. \cite{ilcbm8}
where $\mu =110$ GeV and $\Delta_{EW}=16$.

Formally, these conditions arise from requiring the {\it electroweak fine-tuning measure} 
$\Delta_{EW}$ be not too large, where 
\be
\Delta_{EW} \equiv max_i \left|C_i\right|/(m_Z^2/2)\;,
\ee
may be constructed, with $C_{H_d}=m_{H_d}^2/(\tan^2\beta -1)$, 
$C_{H_u}=-m_{H_u}^2\tan^2\beta /(\tan^2\beta -1)$ and $C_\mu =-\mu^2$.
Also, $C_{\Sigma_u^u(k)} =-\Sigma_u^u(k)\tan^2\beta /(\tan^2\beta -1)$ and 
$C_{\Sigma_d^d(k)}=\Sigma_d^d(k)/(\tan^2\beta -1)$,
where $k$ labels the various loop contributions included in Eq. \ref{eq:mZs}.

The largest of the radiative corrections comes from the top squark sector $\Sigma_u^u(\tst_{1,2})$.
These radiative corrections can be minimized for large stop mixing from a large trilinear $A_t$
parameter, which also raises up the value of $m_h$ to the 125 GeV regime for 
top squark masses in the 1-4 TeV range\cite{ltr}.

An advantage of $\Delta_{EW}$ is that it is model-independent in the sense that any
model which yields the same weak scale mass spectrum will generate the same value
of $\Delta_{EW}$.

\subsection{Fine-tuning of the Higgs mass}

\subsubsection{SM case:}

An alternative measure of EWFT is to require that the (regularized) divergent radiative corrections 
$\delta m_h^2$ to the squared Higgs mass $m_h^2$ be not too large: say $\delta m_h^2\alt m_h^2$. 

In the SM we have
\be
m_{H_{SM}}^2 = 2\mu^2 +\delta m_{H_{SM}}^2 
\label{eq:mHSM}
\ee
where the tree-level squared mass $2\mu^2$ and the quadratically divergent radiative corrections 
\be
\delta m_{H_{SM}}^2\simeq \frac{3}{4\pi^2}\left(-\lambda_t^2+\frac{g^2}{4}+\frac{g^2}{8\cos^2\theta_W}+\lambda\right)\Lambda^2 
\ee
are {\it independent} (here, $\lambda_t$ is the SM top Yukawa coupling, $g$ is the $SU(2)_L$ gauge coupling,
$\lambda$ is the SM Higgs quartic coupling and $\Lambda$ is the effective theory energy cutoff scale). 
Thus, by the EWFT Rule, this is a legitimate fine-tuning evaluation. 
For large $\Lambda$, the large radiative corrections
must be balanced by a fine-tuning of $2\mu^2$ such that $m_{H_{SM}}^2$ maintains its physical value.
Alternatively, to maintain naturalness, then $\delta m_{H_{SM}}^2\sim m_{H_{SM}}^2$ which requires $\Lambda\alt 1$ TeV, 
{\it i.e.} the SM is only valid below about the $\Lambda\sim 1$ TeV scale.

\subsubsection{MSSM case:}

In the MSSM, it is found that 
\be
m_h^2\simeq \mu^2 + m_{H_u}^2+\delta m_{H_u}^2
\ee
where now $\mu^2$ is the {\it supersymmetric} Higgs/higgsino bilinear term which gives
mass to both SM particles (the gauge and Higgs bosons) and the SUSY partner higgsinos.
In addition, $m_{H_u}^2$ is the soft SUSY breaking (SSB) up-Higgs mass term.
If we assume the MSSM is valid up to the GUT scale, then the value of $\delta m_{H_u}^2$ can be found by 
integrating the renormalization group equation (RGE)\cite{bbo}:
\be
\frac{dm_{H_u}^2}{dt}=\frac{1}{8\pi^2}\left(-\frac{3}{5}g_1^2M_1^2-3g_2^2M_2^2+\frac{3}{10}g_1^2 S+3f_t^2 X_t\right)
\label{eq:mHu}
\ee
where $t=\ln (Q^2/Q_0^2)$,
$S=m_{H_u}^2-m_{H_d}^2+Tr\left[{\bf m}_Q^2-{\bf m}_L^2-2{\bf m}_U^2+{\bf m}_D^2+{\bf m}_E^2\right]$
and where $X_t=m_{Q_3}^2+m_{U_3}^2+m_{H_u}^2+A_t^2$.
By neglecting gauge terms and $S$ ($S=0$ in models with scalar soft term universality but can be large in models with non-universality),
and also neglecting the $m_{H_u}^2$ contribution to $X_t$ and the fact that $f_t$ and the soft terms
evolve under $Q^2$ variation,
then this expression may be readily integrated from $m_{SUSY}$ to the cutoff $\Lambda$ to obtain
\be
\delta m_{H_u}^2 \sim -\frac{3f_t^2}{8\pi^2}(m_{Q_3}^2+m_{U_3}^2+A_t^2)\ln\left(\Lambda^2/m_{SUSY}^2 \right) .
\label{eq:DBoE}
\ee
Here, $\Lambda$ may be taken as high as $m_{GUT}\simeq 2\times 10^{16}$ GeV
or even the reduced Planck mass $m_P\simeq 2.4\times 10^{18}$ GeV.
Also, we take $m_{SUSY}^2 \simeq m_{\tst_1}m_{\tst_2}$.
By requiring\cite{kn,papucci,brust,Evans:2013jna} $\Delta_{HS}\sim \delta m_{H_u}^2/(m_h^2/2)\alt 10$ 
one then expects $m_{\tst_{1,2},\tb_1}\alt 600$ GeV.
Using the $\Delta_{HS}$ measure along with $m_h\simeq 125$ GeV then one finds some popular SUSY 
models to be electroweak fine-tuned to 0.1\%\cite{comp}.

Two pitfalls occur within this approach, which are {\it different} from the case of the SM.
\bi
\item The first is that $m_{H_u}^2(\Lambda )$ and $\delta m_{H_u}^2$ are {\it not} independent:
the value of $m_{H_u}^2$ feeds directly into evaluation of $\delta m_{H_u}^2$ via the $X_t$ term.
It also feeds indirectly into $\delta m_{H_u}^2$ by contributing to the evolution of the
$m_{Q_3}^2$ and $m_{U_3}^2$ terms. In fact, the larger the value of $m_{H_u}^2(\Lambda )$, then the
larger is the cancelling correction $\delta m_{H_u}^2$.
Thus, this fine-tuning measure fails under the Fine-tuning Rule.
\item The second is that whereas $SU(2)_L\times U(1)_Y$ gauge symmetry can be broken at tree
level in the SM, in the SUGRA case where SUSY is broken in a hidden sector via the superHiggs mechanism 
then $m_{H_u}^2\sim m_{3/2}^2>0$ and EW symmetry is not even broken until
one includes radiative corrections. For SUSY models valid up to some high scale $\Lambda\gg m_{weak}$,
EW symmetry is broken radiatively by $m_{H_u}^2$ being driven to large negative values 
by the large top quark Yukawa coupling\cite{rewsb}.
\ei
By combining dependent terms, then we have a regrouping\cite{ltr,rns}
\be
m_h^2|_{phys}=\mu^2+\left(m_{H_u}^2(\Lambda )+\delta m_{H_u}^2 \right)
\label{eq:mh}
\ee
where now $\mu^2$ and $\left(m_{H_u}^2(\Lambda )+\delta m_{H_u}^2 \right)$ are each independent so each 
should be comparable to $m_h^2$ in order to avoid fine-tuning. 
It is often claimed that under such a regrouping, then the SM Higgs mass would also not be fine-tuned. 
But here we see that in the MSSM case-- since the $m_{H_u}^2$ and $\delta m_{H_u}^2$ terms are dependent-- 
the situation is different from the SM and one must lump dependent terms together.
The regrouping in Eq. \ref{eq:mh} of contributions to $m_h^2$ into independent terms leads back to the $\Delta_{EW}$ measure.

\subsection{$\Delta_{BG}$ and model-dependence}

The more traditional measure $\Delta_{BG}$ was proposed by Ellis {\it et al.}\cite{eenz} 
and later investigated more thoroughly by Barbieri and Giudice\cite{bg}. 
The starting point is to express $m_Z^2$ in terms of weak scale SUSY parameters
as in Eq. \ref{eq:mZs}:
\be
m_Z^2 \simeq -2m_{H_u}^2-2\mu^2
\label{eq:mZsapprox}
\ee
where the partial equality obtains for moderate-to-large $\tan\beta$ values and where we assume for
now the radiative corrections are small.
An advantage of $\Delta_{BG}$ over the previous large-log measure is that it maintains the 
correlation between $m_{H_u}^2(\Lambda )$ and $\delta m_{H_u}^2$ by replacing
$m_{H_u}^2 (m_{weak})= \left( m_{H_u}^2(\Lambda )+\delta m_{H_u}^2\right)$ by its expression in 
terms of high scale parameters.
To evaluate $\Delta_{BG}$, one needs to know the explicit dependence of $m_{H_u}^2$ and $\mu^2$ on the
fundamental parameters. 
Semi-analytic solutions to the one-loop renormalization group equations
for $m_{H_u}^2$ and $\mu^2$ can be found for instance in Ref's \cite{munoz}.
For the case of $\tan\beta =10$, then\cite{abe,martin,feng}
\bea
m_Z^2& \simeq & -2.18\mu^2 + 3.84 M_3^2+0.32M_3M_2+0.047 M_1M_3-0.42 M_2^2 \nonumber \\
& & +0.011 M_2M_1-0.012M_1^2-0.65 M_3A_t-0.15 M_2A_t\nonumber \\
& &-0.025M_1 A_t+0.22A_t^2+0.004 M_3A_b\nonumber \\
& &-1.27 m_{H_u}^2 -0.053 m_{H_d}^2\nonumber \\
& &+0.73 m_{Q_3}^2+0.57 m_{U_3}^2+0.049 m_{D_3}^2-0.052 m_{L_3}^2+0.053 m_{E_3}^2\nonumber \\
& &+0.051 m_{Q_2}^2-0.11 m_{U_2}^2+0.051 m_{D_2}^2-0.052 m_{L_2}^2+0.053 m_{E_2}^2\nonumber \\
& &+0.051 m_{Q_1}^2-0.11 m_{U_1}^2+0.051 m_{D_1}^2-0.052 m_{L_1}^2+0.053 m_{E_1}^2 ,
\label{eq:mZsparam}
\eea
where all terms on the right-hand-side are understood to be $GUT$ scale parameters.

Then, the proposal is that the variation in $m_Z^2$ with respect to
parameter variation be small:
\be
\Delta_{BG}\equiv max_i\left[ c_i\right]\ \ {\rm where}\ \ 
c_i=\left|\frac{\partial\ln m_Z^2}{\partial\ln a_i}\right|
=\left|\frac{a_i}{m_Z^2}\frac{\partial m_Z^2}{\partial a_i}\right|
\label{eq:DBG}
\ee
where the $a_i$ constitute the fundamental parameters of the model.
Thus, $\Delta_{BG}$ measures the fractional change in $m_Z^2$ due to fractional variation in
high scale parameters $a_i$.
The $c_i$ are known as {\it sensitivity coefficients}\cite{feng}. 

The requirement of low $\Delta_{BG}$ is then equivalent to the requirement of no 
large cancellations on the right-hand-side of Eq. \ref{eq:mZsparam} since (for linear terms) 
the logarithmic derivative just picks off coefficients of the relevant parameter. For instance, 
$c_{m_{Q_3}^2}=0.73\cdot (m_{Q_3}^2/m_Z^2)$. If one allows $m_{Q_3}\sim 3$ TeV (in accord with 
requirements from the measured value of $m_h$) then one obtains $c_{m_{Q_3}^2}\sim 800$
and so $\Delta_{BG}\ge 800$. In this case, SUSY would be electroweak fine-tuned to about 0.1\%. 
If instead one sets $m_{Q_3}=m_{U_3}=m_{H_u}\equiv m_0$ as in models with scalar mass universality, then the various
scalar mass contributions to $m_Z^2$ largely cancel and $c_{m_0^2}\sim -0.017 m_0^2/m_Z^2$: 
the contribution to $\Delta_{BG}$ from scalars drops by a factor $\sim 50$. 

The above argument illustrates the extreme model-dependence of $\Delta_{BG}$ for multi-parameter SUSY models. 
The value of $\Delta_{BG}$ can change radically from theory to theory even if those theories 
generate exactly the same weak scale sparticle mass spectrum. The model dependence of $\Delta_{BG}$
arises due to a violation of the Fine-tuning Rule: one must combine dependent terms into independent quantities
before evaluating EW fine-tuning.

\subsection{When is $\Delta_{BG}$ a reliable measure of naturalness?}

In Ref. \cite{comp}, it was argued that in an ultimate theory (UTH), where all soft parameters are correlated, 
then $\Delta_{BG}$ should be a reliable measure of naturalness. In fact, most supersymmetric theories with SUSY breaking
generated in a hidden sector fulfill this requirement. For instance, in supergravity theories with hidden sector
SUSY breaking via the superHiggs mechanism, then all soft breaking parameters are expected to be some multiple 
of the gravitino mass $m_{3/2}$. (For example, in string theory with dilaton-dominated SUSY breaking\cite{kl,Brignole:1993dj}, 
then we expect $m_0^2=m_{3/2}^2$ with $m_{1/2}=-A_0=\sqrt{3}m_{3/2}$).
For any fully specified hidden sector, we expect each SSB term to be some multiple of $m_{3/2}$: {\it e.g.}
\bea
m_{H_u}^2&=&a_{H_u}\cdot m_{3/2}^2,\label{eq:1} \\
m_{Q_3}^2&=&a_{Q_3}\cdot m_{3/2}^2,\\
A_t&=&a_{A_t}\cdot m_{3/2},\\
M_i&=&a_i\cdot m_{3/2},\\
& & \cdots \label{eq:5} .
\eea
Here, the coefficients $a_i$ {\it parametrize our ignorance} of the exact model for SUSY breaking.
By using several adjustable parameters, we cast a wide net which encompasses a large range of hidden sector SUSY breaking possibilities. 
But this doesn't mean that each SSB parameter is expected to be independent of the others. 
It just means we do not know how SUSY breaking occurs, and how the soft terms are correlated:
it is important not to confuse parameters which ought to be related to one another 
in any sensible theory of SUSY breaking with independently adjustable soft SUSY breaking terms.

Now, plugging the soft terms \ref{eq:1}-\ref{eq:5} into Eq. \ref{eq:mZsparam}, one arrives at the expression
\be
m_Z^2=-2.18\mu^2 +a\cdot m_{3/2}^2 .
\ee
The value of $a$ is just some number which is the sum of all the coefficients of the terms $\propto m_{3/2}^2$.
For now, we assume $\mu$ is independent of $m_{3/2}$ as will be discussed shortly.

In this case, we can compute the sensitivity coefficients:\footnote{In mAMSB, the soft terms are 
also written as multiples of $m_{3/2}$ or $m_{3/2}^2$. In mGMSB, the soft terms are written as multiples
of messenger scale $\Lambda_m$. The argument proceeds in an identical fashion in these cases.}
\bea
c_{m_{3/2}^2} &=& |a\cdot (m_{3/2}^2/m_Z^2)|\ \ {\rm and}\label{eq:A} \\
c_{\mu^2} &=& |-2.18 (\mu^2/m_Z^2 )|.\label{eq:B}
\eea
For $\Delta_{BG}$ to be $\sim 1-10$ (natural SUSY with low fine-tuning), then Eq. \ref{eq:B} implies
\bi
\item $\mu^2 \sim m_Z^2$ 
\ei
and also Eq. \ref{eq:A} implies 
\bi
\item $a\cdot m_{3/2}^2\sim m_Z^2$.
\ei 
The first of these conditions implies light higgsinos with mass $\sim 100-200$ GeV, the closer to $m_Z$ 
the better. The second condition can be satisfied if $m_{3/2}\sim m_Z$\cite{bg} (which now seems highly unlikely
due to a lack of LHC8 SUSY signal\footnote{For instance, in simple SUGRA models, then the 
scalar masses $m_0=m_{3/2}$. Since LHC requires rather high $m_0$, then we would also expect rather large $m_{3/2}$.} 
and the rather large value of $m_h$) {\it or} if $a$ is quite small:
in this latter case, the SUSY soft terms conspire such that there are large cancellations amongst the various
coefficients of $m_{3/2}^2$ in Eq. \ref{eq:mZsparam}: 
this is what is called radiatively-driven natural SUSY\cite{ltr,rns} since in this case a large high scale 
value of $m_{H_u}^2$ can be driven radiatively to small values $\sim -m_Z^2$ at the weak scale.

Furthermore, we can equate the value of $m_Z^2$ in terms of weak scale parameters with the value 
of $m_Z^2$ in terms of GUT scale parameters:
\be
m_Z^2\simeq -2\mu^2(weak)-2m_{H_u}^2(weak) \simeq -2.18\mu^2(GUT)+a\cdot m_{3/2}^2 .
\ee
Since $\mu$ hardly evolves under RG running (the factor 2.18 is nearly 2), then we have the
BG condition for low fine-tuning as 
\be
-m_{H_u}^2(weak) \sim a\cdot m_{3/2}^2\sim m_Z^2 ,
\ee
{\it i.e.} that the value of $m_{H_u}^2$ must be driven to 
{\it small} negative values $\sim - m_Z^2$ at the weak scale. 
These are exactly the conditions required by the model-independent EWFT
measure $\Delta_{EW}$: {\it i.e.} we have
\be
\lim_{n_{SSB}\to 1} \Delta_{BG}\to \Delta_{EW}
\ee
where $n_{SSB}$ is the number of {\it independent} soft SUSY breaking terms.
Of course, this approach also reconciles the Higgs mass fine-tuning measure (with
appropriately regrouped independent terms) with the $\Delta_{BG}$ measure 
(when applied to models with a single independent soft breaking term such as $m_{3/2}$).
%
%

\subsection{The $\mu$ parameter and some solutions to the $\mu$ problem}

One of the central problems of supersymmetric theories concerns the origin of the
superpotential $\mu$ term: $W\ni \mu H_u H_d$. Since this term is supersymmetric 
(does not arise from SUSY breaking) its value might be expected to be $\mu\sim M_P$.
However, phenomenology dictates instead that $\mu\sim m_{weak}$. A variety of solutions to 
the SUSY $\mu$ problem arise in the literature. Here we comment briefly on three of them.

\subsubsection{NMSSM}

In the Next-to-Minimal Supersymmetric Standard Model (NMSSM)\cite{nmssm}, one assumes some
symmetry forbids the usual $\mu$ term, but then a visible sector gauge singlet superfield $S$ is 
added with superpotential
\be
W\ni \lambda_S SH_uH_d . 
\ee
The scalar component of $S$ develops a vev $\langle S\rangle\sim m_{3/2}$ which generates a $\mu$
term: $\mu =\lambda_S  \langle S\rangle\sim m_{3/2}$. In addition to the $\mu$ term, one obtains 
new physical Higgs particles along with a singlino. An additional contribution to the Higgs mass
is also generated which some authors find appealing. 

A drawback to this scenario is that introduction of true gauge singlets may lead back to 
destabilizing the gauge hierarchy via tadpole diagrams\cite{ell,bp}. This destabilization can be avoided
by introducing $S$ as a composite object\cite{rr} although this leads to possibly recondite models
and a movement away from parsimony.

\subsubsection{Giudice-Masiero}

In the Giudice-Masiero (GM) mechanism\cite{GM}, it is assumed that the usual $\mu$ term is forbidden by
some symmetry which is applicable to the visible sector but which is not respected by hidden sector fields.
In such a case, then there may exist a (non-renormalizable) coupling of Higgs doublets to the hidden sector
such as 
\be
K\ni \lambda h_m H_u H_d /M_P
\ee
where $h_m$ is a hidden sector field. When $h_m$ develops a SUSY breaking vev $\langle F_h\rangle\sim m_s^2$ 
where $m_s$ is the hidden sector mass scale (with $m_{3/2}\sim m_s^2/M_P$), then a $\mu$ term is generated with 
\be
\mu\sim\lambda \langle F_h\rangle/M_P\sim \lambda m_{3/2}   .
\ee
Thus, in the GM solution, we expect $\mu\sim m_{3/2}$. If we expect $m_{3/2}\gg 1$ TeV scale due to 
lack of LHC signal, then we would arrive at high EW fine-tuning unless $\lambda$ was tiny.\footnote{
In a recent paper\cite{Leggett:2014mza}, the authors argue that no-scale SUSY models contain
only one free parameter $m_{3/2}$, and where $\mu\sim m_{3/2}$ is generated via GM mechanism so that
$m_Z^2=a\cdot m_{3/2}^2$ where $a$ is some constant. In such a case, it is a tautology that 
$\Delta_{BG}=c_{m_{3/2}}=|\partial\ln m_Z^2/\partial\ln m_{3/2}^2 |=1$ and there is no fine-tuning. 
However, in this case the authors do not
produce an explicit hidden sector-visible sector coupling which produces exactly the right $\mu$
value which is required to generate $m_Z=91.2$ GeV. 
In the absence of an explicit hidden sector model, then one must regard
instead $\mu$ as a free parameter which parametrizes our ignorance of the hidden sector, so that
there are actually two free parameters with $m_Z^2\sim -2.18 \mu^2 +a\cdot m_{3/2}^2$. 
Then as usual large $\mu$ will require high fine-tuning.
}

\subsubsection{Kim-Nilles}

The Kim-Nilles (KN) mechanism\cite{KN} arises as a byproduct of the PQ solution to the strong CP problem
and is the supersymmetric extension of the DFSZ axion model\cite{dfsz}. 
In KN, the $H_u$ and $H_d$ superfields carry PQ charges $Q_u$ and $Q_d$ so the usual $\mu$ term 
is forbidden by PQ symmetry. An additional visible sector field $P$ carrying PQ charge $-(Q_u+Q_d)/2$ 
is then required so that the superpotential term
\be
W_{DFSZ}\ni \lambda P^2 H_u H_d/M_P
\ee
is present. The PQ symmetry is broken, for instance, by a superpotential\cite{chun_dfsz}
\be
W_{PQ}\ni \lambda_S S\left( PQ-f_a^2 \right)
\ee
(the PQ charge of $Q$ and $S$ is $-Q_P$ and 0 respectively)
which leads to $\langle P\rangle\sim \langle Q\rangle\sim f_a$. 
The axion-axino-saxion fields are combinations of the $P$ and $Q$ fields. 
A $\mu$ term is then generated with
\be
\mu\sim \lambda f_a^2/M_P .
\ee
Originally, Kim-Nilles had sought to identify the PQ scale $f_a$ with the hidden sector SUSY breaking scale $m$.
However, now we see that in fact the developing Little Hierarchy $\mu\ll m_{3/2}$ is 
nothing more than a reflection of an apparent mis-match between the PQ breaking scale and 
hidden sector SUSY breaking scale $f_a\ll m_s$.
Guided by electroweak naturalness, we expect $\mu\sim 100-200$ GeV so that with $\lambda\sim 1$, then
we expect 
\be
f_a\sim 10^{10}\ {\rm GeV} .
\ee
In this case, since the axion mass 
$m_a\sim 6.2\ \mu {\rm eV}\left( \frac{10^{12}\ {\rm GeV}}{f_a}\right)$
then {\it the Higgs mass tells us where to look for the axion}: $m_a\sim 620\mu{\rm eV}$ with DFSZ couplings.
Furthermore, in such a scenario then one expects dark matter to consist of a DFSZ axion along with 
a higgsino-like WIMP: {\it i.e. two} dark matter particles\cite{dfsz1}.

\section{Numerical procedure}

In the following Section, we will evaluate EW fine-tuning for a variety of SUSY models using the 
case where all three measures agree since as shown above the
Higgs mass fine-tuning and the BG measure both reduce to $\Delta_{EW}$ once dependent contributions to
$m_Z^2$ or $m_h^2$ are combined into independent terms.

For each model, we generate random sets of parameter values over the range 
listed in each subsection, and then generate supersymmetric sparticle and Higgs 
mass spectra using the Isasugra\cite{isasugra} subprogram of Isajet\cite{isajet}.
We require of each solution that:
\bi
 \item electroweak symmetry be radiatively broken (REWSB),
 \item the neutralino $\tz_1$ is the lightest MSSM particle,
 \item the light chargino mass obeys the model independent LEP2 limit 
$m_{\tw_1}>103.5$~GeV\cite{lep2ino} ($m_{\tw_1}>91.9$ GeV in the case of a wino-like chargino) and
\item $m_h= 125.5\pm 2.5$~GeV.
\ei
We do not impose any LHC sparticle search limits since our general scan can
produce compressed spectra which in many cases can easily elude LHC gluino and squark searches.
We also do not impose WIMP dark matter constraints since for cases with a standard thermal 
WIMP underabundance, the WIMP abundance might be augmented by late decaying cosmological
relics ({\it e.g.} axinos, saxions, moduli, $\cdots$) or in the case of an overabundance, the
WIMPs might decay to yet lighter particles ({\it e.g.} into light axino LSPs) or be diluted
by late time entropy injection\cite{kimreview}.

We will also calculate the values of $BF(b\to s\gamma )$\cite{vb_bsg,isabsg} and 
$BF(B_S\to\mu^+\mu^- )$\cite{isabmm} for each point generated (we also calculate
other $B$ decay observables which turn out to be far less constraining).
The measured value of $BF(b\to s\gamma )$ is found to be $(3.55\pm 0.26)\times 10^{-4}$~\cite{Asner:2010qj}.
For comparison, the SM prediction\cite{Misiak:2006zs} is $BF^{SM}(b\to s\gamma )=(3.15\pm 0.23)\times 10^{-4}$.
Also, recently both the LHCb collaboration\cite{lhcb} and CMS\cite{cms_Bs} have 
measured events interpretted as $B_s\to\mu^+\mu^-$. 
Their combined branching fraction is determined to be $BF(B_s\to\mu^+\mu^- )=(2.9\pm 0.7)\times 10^{-9}$
which is in rough accord with the SM prediction of $(3.2\pm 0.2)\times 10^{-9}$. 
Here, SUSY model points with 
\bi
\item $BF(b\to s\gamma )= (3.03-4.08)\times 10^{-4}$ 
\ei
and 
\bi
\item $BF(B_s\to\mu^+\mu^- )=(1.5-4.3)\times 10^{-9}$ 
\ei 
will be labeled as satisfying $B$-physics constraints.

\section{Electroweak fine-tuning in various SUSY models}

\subsection{mSUGRA/CMSSM}

First we scan over the paradigm mSUGRA\cite{msugra} or CMSSM\cite{kane} model with parameter ranges given by
\bi
\item $m_0:0-15$ TeV,
\item $m_{1/2}: 0-2$ TeV,
\item $-2.5<A_0/m_0<2.5:$
\item $\tan\beta :3-60$,
\ei
and for both signs of $\mu$.\footnote{Our convention for $\mu$ gives a positive contribution to 
$(g-2)_\mu$ when $\mu >0$.} The results of this scan have been shown previously in 
Ref. \cite{dewsug} for all $\tan\beta$ and in Ref. \cite{comp} for $\tan\beta =10$. 
We present it here for completeness
so that the reader may more readily compare these results against other SUSY models, and because now we also
impose more restrictive $B$-decay constraints.

\begin{figure}[tbp]
\includegraphics[height=0.4\textheight]{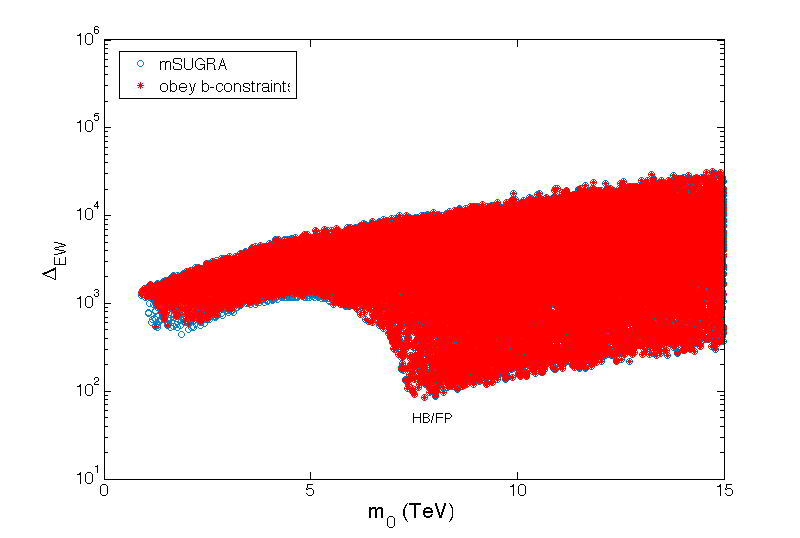}
\caption{Plot of $\Delta_{EW}$ vs. $m_0$ from a scan over 
mSUGRA/CMSSM parameters space whilst maintaining $m_h=125.5\pm 2.5$ GeV 
and whilst obeying $B$-decay constraints.
The location of the hyperbolic branch/focus point regions is labelled as HB/FP.
\label{fig:msugra}}
\end{figure}

The value of $\Delta_{EW}$ is shown vs. $m_0$ in Fig. \ref{fig:msugra} where blue dots comprise all
solutions while red dots also respect $B$-decay constraints. 
For low $m_0$, the value of $\Delta_{EW}$ is around $10^3$, indicating
EWFT at the $\Delta_{EW}^{-1}\sim 0.1\%$ level. As $m_0$ increases, the value of $\Delta_{EW}$ can drop
sharply into the $10^2$ range for $m_0\sim 7-10$ TeV. This is the case of the 
hyperbolic branch/focus-point region (denoted HB/FP) where $\mu$ becomes small\cite{ccn,fp}. 
The value of $\Delta_{EW}$ doesn't drop to arbitrarily small values
because at such large $m_0$ values then the top squark masses become $\sim 5-10$ TeV and the radiative 
corrections $\Sigma_u^u(\tst_{1,2} )$ become large. In fact, as $m_0$ increases beyond 7 TeV, then
the minimum of $\Delta_{EW}$ also increases due to the increasing radiative corrections.
With such a high minimum value of $\Delta_{EW}$, we would expect mSUGRA/CMSSM probably does 
not describe nature.

\subsection{NUHM1}

The NUHM1 model\cite{nuhm1} is inspired by $SO(10)$ SUSY GUT models where the Higgs doublets 
live in the 10-dimensional fundamental representation while the matter scalars inhabit the 16-dimensional 
spinor representation. 
In this case, the parameter set is expanded by one and now we scan over 
\bi
\item $m_0:0-15$ TeV,
\item $m_{H_u}=m_{H_d}\equiv m_H:0-15$ TeV,
\item $m_{1/2}: 0-2$ TeV,
\item $-2.5<A_0/m_0<2.5:$
\item $\tan\beta :3-60$.
\ei
By increasing $m_H\gg m_0$, then $m_{H_u}^2$ is only driven to small instead of large negative values, 
while if $m_{H_u}^2$ is increased too much, then $m_{H_u}^2$ is never driven negative and 
electroweak symmetry is not broken. 
If $m_{H}$ is taken smaller than $m_0$, even with $m_{H}^2<0$ as a possibility, then $m_{H_d}\sim m_A$ can be 
decreased while $m_{H_u}^2$ is driven to very large negative values. 
In the former case, where $m_{H_u}^2$ is driven to small negative values, 
then $\mu$ also decreases-- since its value is set to yield the measured $Z$ mass via Eq. \ref{eq:mZs}. 
In such cases, we expect reduced values of $\Delta_{EW}$. 

\begin{figure}[tbp]
\includegraphics[height=0.4\textheight]{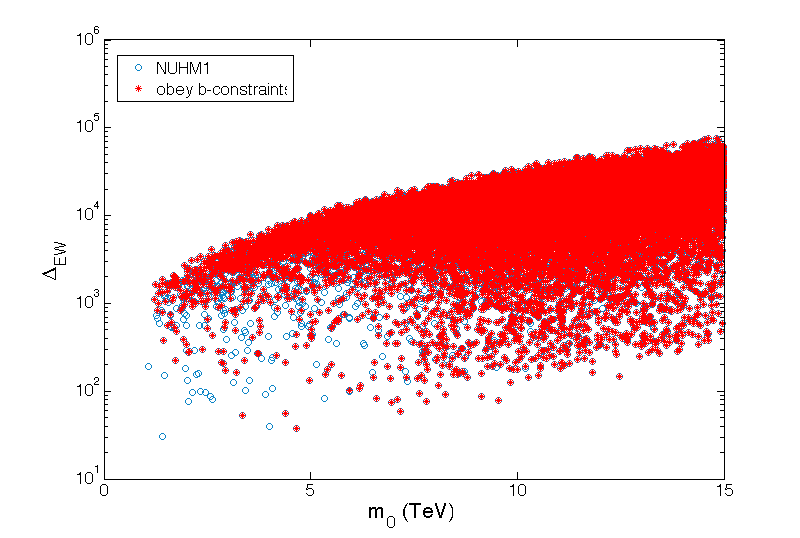}
\caption{Plot of $\Delta_{EW}$ vs. $m_0$ from a scan over NUHM1 parameters 
space whilst maintaining $m_h=125.5\pm 2.5$ GeV.
\label{fig:nuhm1}}
\end{figure}

In the scan results shown in Fig. \ref{fig:nuhm1}, this is indeed bourne out, as we see that
the minimal value of $\Delta_{EW}$ reaches as low as $\sim 30$, which is much less fine-tuned than mSUGRA.
Values of $\Delta_{EW}$ in the $30-50$ range which obey $B$-decay constraints and $m_h\sim 125$ can be found
for $m_0\sim 3-10$ TeV. With such large $m_0$ values, then the top squarks also tend to be in the
$3-10$ TeV regime and the top squark radiative corrections prevent $\Delta_{EW}$ from reaching
below $\sim 30$.

\subsection{NUHM2}

The NUHM2 model\cite{nuhm2} is inspired by $SU(5)$ SUSY GUTs where each of the MSSM Higgs doublets live in
separate 5 and $\overline{5}$ representations, or by $SO(10)$ SUSY GUTs with $D$-term scalar 
mass splitting. In such a case, we expand the mSUGRA parameter space
to include $m_{H_u}^2$ and $m_{H_d}^2$ as soft SUSY breaking terms which are independent of $m_0$.
Using weak scale mass relations, then $m_{H_u}^2$ and $m_{H_d}^2$ can be traded for the more
convenient weak scale parameters $\mu$ and $m_A$.

\begin{figure}[tbp]
\includegraphics[height=0.4\textheight]{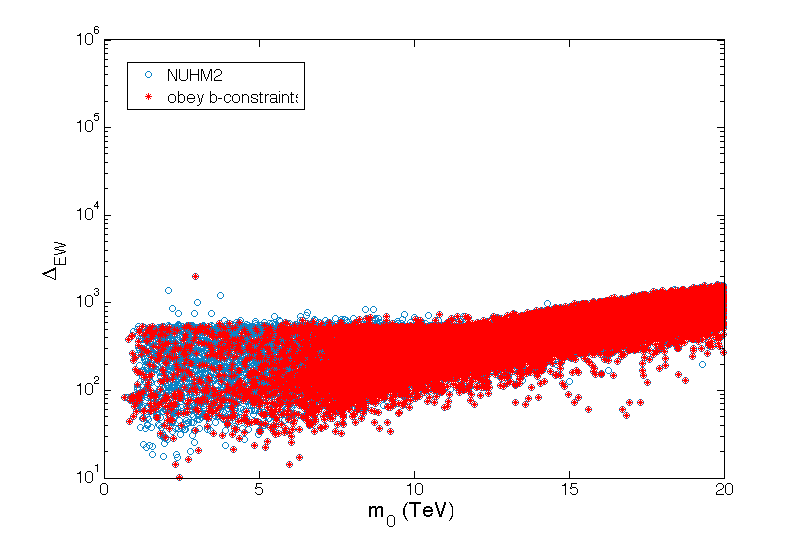}
\caption{Plot of $\Delta_{EW}$ vs. $m_0$ from a scan over NUHM2 parameters 
space whilst maintaining $m_h=125.5\pm 2.5$ GeV.
\label{fig:nuhm2}}
\end{figure}

In the case of NUHM2, we scan in accord with Ref. \cite{rns}
\bi
\item $m_0:\ 0-20$ TeV,
\item $m_{1/2}:\ 0.3-2$ TeV,
\item $-3<\ A_0/m_0<3$
\item $\mu:\ 0.1-1.5$ TeV,
\item $m_A:\ 0.15-1.5$ TeV,
\item $\tan\beta :3-60$,
\ei
with results shown in Fig. \ref{fig:nuhm2}. Here, we see that $\Delta_{EW}$ can reach values
as low as 10, corresponding to $\Delta_{EW}^{-1}\sim 10\%$ EWFT. Even lower values $\sim 7$ have been 
generated in Ref. \cite{comp} for a fixed $\tan\beta =10$ value. The key here is that low $\mu$
values $\sim 100-200$ GeV can be input by hand while top squarks can occur in the $1-5$ TeV
regime with large mixing, which also acts to reduce the radiative corrections 
$\Sigma_u^u(\tst_{1,2})$\cite{ltr}. 
The required GUT scale values of $m_{H_u}$ are about $1.2 m_0$ while $m_{H_d}(m_{GUT})$ can be anywhere in the TeV-range\cite{rns}.
As $m_0$ increases beyond about 7 TeV, then the min of
$\Delta_{EW}$ slowly increases due to increasing top squark radiative corrections.
For the model examined in Ref. \cite{rns} with split generations, then $2-4$ TeV top squarks 
are allowed in accord  with $10-30$ TeV first/second generation scalars: this situation offers at least
a partial decoupling solution to the SUSY flavor and CP problems\cite{dine}.

\subsection{mGMSB}

In minimal GMSB\cite{gmsb,Draper:2011aa}, a sector of ``messenger'' fields is hypothesized which communicates 
between the hidden SUSY breaking sector and the visible/MSSM sector. 
Visible sector scalar fields acquire a mass $m_i^2\propto (\alpha_i/4\pi)^2\Lambda^2$ while gauginos acquire a mass
$M_i=(\alpha_i/4\pi)\Lambda$ where $\Lambda$ parametrizes the induced SUSY breaking scale in the messenger sector.
The trilinear SSB $a$-terms are suppressed by an additional loop factor and hence are expected to be small.
This latter effect leads to only small amounts of stop mixing: consequently  huge stop masses are required in mGMSB
in order to generate $m_h\sim 125$ GeV. Furthermore, the hierarchy of mass values in mGMSB
\be
M_1<m_E<M_2<m_L=m_{H_u}=m_{H_d}\ll M_3<m_{\tq}
\ee
means that the $m_{H_u}^2$ boundary condition is already suppressed at 
the messenger scale $M_{mes}$, and then is strongly driven to large negative values due to the large values of
$m_{Q_3}$ and $m_{U_3}$ contributing to the $X_t$ term in Eq. \ref{eq:mHu}. 
The upshot is that for allowed parameter ranges, $m_{H_u}^2$ is driven to large negative values at the weak scale, 
and the value of $\mu$ must be large positive (fine-tuned) to obtain the measured value of $m_Z$. 

\begin{figure}[tbp]
\includegraphics[height=0.4\textheight]{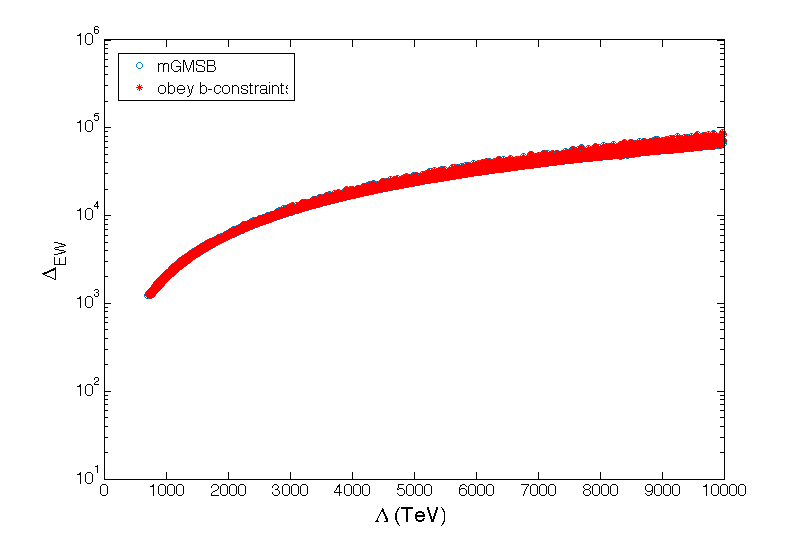}
\caption{Plot of $\Delta_{EW}$ vs. $\Lambda$ from a scan over mGMSB parameters 
space whilst maintaining $m_h=125.5\pm 2.5$ GeV.
\label{fig:gmsb}}
\end{figure}

Our results are shown in Fig. \ref{fig:gmsb} where we plot $\Delta_{EW}$ vs. $\Lambda$ from a scan over values
\bi
\item $\Lambda:\ 10^2-10^4$ TeV,
\item $M_{mes}=2\Lambda$,
\item $\tan\beta :3-60$,
\item $sign (\mu )=\pm$.
\ei
From the plot, we see that requiring $m_h:123-128$ GeV then requires $\Lambda\agt 500$ TeV, 
which results in very heavy top squarks and large fine-tuning, with the minimum of $\Delta_{EW}$ at $10^3$, or 0.1\% EWFT.
Here, we would conclude that at least minimal GMSB is not likely to describe nature. 
%

\subsection{mAMSB}

In anomaly-mediated SUSY breaking models\cite{amsb}, 
it is assumed that the SUSY breaking sector is sequestered
from the visible sector-- perhaps in extra spacetime dimensions-- so that the dominant 
soft SUSY breaking contribution comes from the superconformal anomaly. In this case, 
gaugino masses $M_i= c_i (g_i^2/16\pi^2)  m_{3/2}$ with $c_i=(33/5,1,-3)$ for the $U(1)$, 
$SU(2)$ and $SU(3)$ groups respectively. Thus, multi-TeV values of $m_{3/2}$ are required
which also ameliorates the so-called cosmological gravitino problem\cite{gravprob}. 
Also, the lightest gauginos are
wino-like with a neutral wino as LSP. Due to tachyonic slepton masses in pure AMSB, an additional
universal contribution $m_0^2$ is invoked in order to gain a phenomenologically viable
spectrum of matter scalars. Since the trilinear $a$ parameter is small, mAMSB has trouble
generating $m_h\sim 125$ GeV unless top squarks are in the multi-TeV regime.

\begin{figure}[tbp]
\includegraphics[height=0.4\textheight]{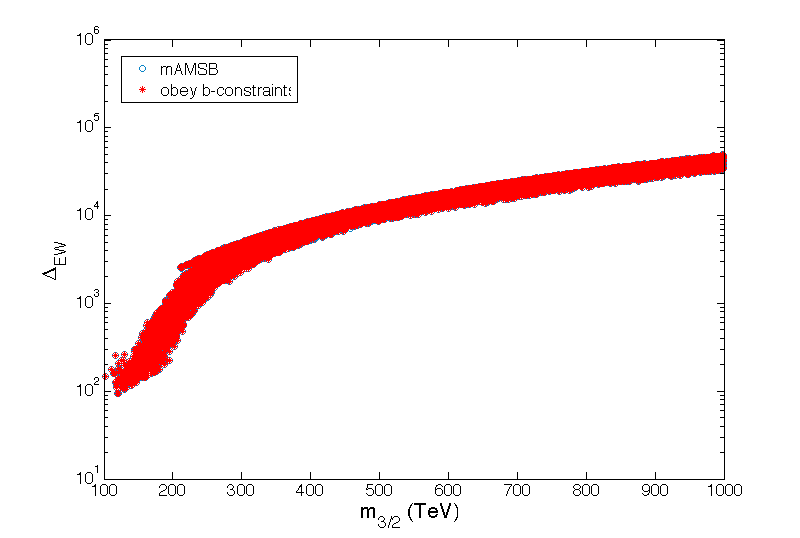}
\caption{Plot of $\Delta_{EW}$ vs. $m_{3/2}$ from a scan over mAMSB parameters 
space whilst maintaining $m_h=125.5\pm 2.5$ GeV.
\label{fig:mamsb}}
\end{figure}

We scan over mAMSB parameter space according to 
\bi
\item $m_{3/2}:\ 20-1000$ TeV,
\item $m_0:\ 0-10$ TeV,
\item $\tan\beta :3-60$,
\item $sign (\mu )=\pm$.
\ei
Our results are shown in Fig. \ref{fig:mamsb}. Here, we see that the minimal value of 
$\Delta_{EW}$ occurs at $m_{3/2}\sim 100$ TeV and has a value $\sim 100$, or 1\% EWFT.  
\subsection{HCAMSB}

An alternative set-up for AMSB, known as hypercharged anomaly-mediation (HCAMSB), 
has been advocated in Ref.~\cite{dermisek}.  It is a string motivated scenario which uses a similar
construction to the one envisioned for AMSB. In HCAMSB, SUSY breaking is localized at the bottom
of a strongly warped hidden region, geometrically separated from the visible region where the
MSSM resides. The warping suppresses contributions due to tree-level gravity mediation[4]
so that anomaly mediation[1] can become the dominant source of SUSY breaking in the visible
sector. Possible exceptions to this sequestering mechanism are gaugino masses of U(1) gauge
symmetries [5]. Thus, in the MSSM, the mass of the bino (the gaugino of $U(1)_Y$) can be the
only soft SUSY breaking parameter not determined by anomaly mediation[6]. Depending on
its size, the bino mass $M_1$ can lead to a small perturbation to the spectrum of anomaly mediation,
or it can be the largest soft SUSY breaking parameter in the visible sector. As a result of
RG evolution, its effect on other soft SUSY breaking parameters can dominate the contribution
from anomaly mediation. In extensions of the MSSM, additional $U(1)′$s can also communicate
SUSY breaking to the MSSM sector [7].

In HCAMSB, the SSB terms are of the same form as AMSB except for the $U(1)_Y$ gaugino mass:
\be
M_1 = \tilde{M}_1 + \frac{b_1g_1^2}{16\pi^2} m_{3/2},
\ee
where $\tilde{M}_1=\alpha m_{3/2}$. 
The large $U(1)_Y$ gaugino mass can cause $m_{H_u}^2$ to first run to large positive values
before it is driven negative so that EW symmetry is broken. This potentially leads to lower
fine-tuning since then $m_{H_u}^2$ may be driven to just small negative values.

\begin{figure}[tbp]
\includegraphics[height=0.4\textheight]{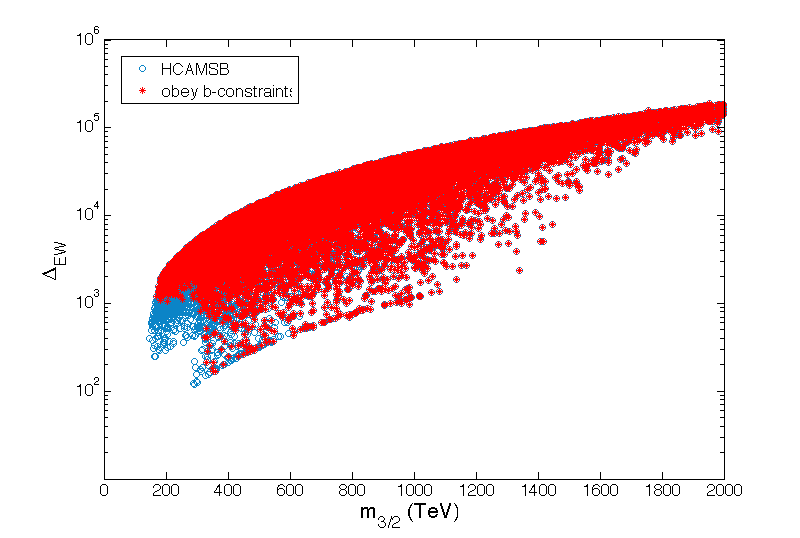}
\caption{Plot of $\Delta_{EW}$ vs. $m_{3/2}$ from a scan over HCAMSB parameters 
space whilst maintaining $m_h=125.5\pm 2.5$ GeV.
\label{fig:hcamsb}}
\end{figure}

We scan over the HCAMSB parameter space
\bi
\item $m_{3/2}:\ 25-2000$ TeV,
\item $\alpha:\ -0.25-0.25$,
\item $\tan\beta :3-60$,
\item $sign (\mu )=\pm$
\ei
with the LEP2 chargino mass limit reduced to $m_{\tw_1}>91.9$ GeV as appropriate for a wino-like LSP.
Our results are shown in Fig. \ref{fig:hcamsb} where we plot $\Delta_{EW}$ vs. $m_{3/2}$.
Here, we find a minimal value of $\Delta_{EW}\sim 100$ for $m_{3/2}\sim 400$ TeV.

\subsection{Mixed moduli-anomaly mediation}

These models, known as mixed moduli-anomaly mediated SUSY breaking (MMAMSB), or mirage mediation, 
 are based on the KKLT construction\cite{kklt} of string compactification with fluxes,
which produce the necessary de Sitter vacuum.
In the KKLT construct, one first introduces nonzero fluxes in the Type IIB string theory 
compactified on a Calabi-Yau manifold. 
Due to the nonzero fluxes, the complex structure moduli and the dilaton are completely fixed but
the size modulus $T$ remains a flat direction. 
To fix this, KKLT invoked non-perturbative effects, such as gaugino condensation on D7 branes. 
At this stage, all moduli are fixed, but one ends up with supersymmetric 
vacua and negative vacuum energy. 
The final step in the construction is to include anti $D$-branes yielding the desired
de-Sitter vacua (with positive vacuum energy) and breaking supersymmetry. Because of
the presence of branes and fluxes, the models have generically warped compactifications.
Due to the warping, the addition of the anti D-brane breaks supersymmetry by a very small amount.

The phenomenology of KKLT-inspired models is distinctive in that moduli fields 
and the Weyl anomaly make comparable contributions to SUSY breaking effects
in the observable sector of fields\cite{mmamsb}. 
The contribution of each can be parametrized by $\alpha$
which yields pure AMSB for $\alpha =0$ but which tends to pure moduli (gravity) mediation as
$\alpha$ becomes large. The phenomenology also depends on the so-called modular weights 
which in turn depend on the location of various fields in the extra dimensions:
$n_i = 0\ (1)$ for matter fields located on D7 (D3) branes; fractional
values $n_i = 1/2$ are also possible for matter living at brane intersections. 
It is claimed that MMAMSB models have the potential to be minimally 
EW fine-tuned\cite{Choi:2006xb,Lebedev:2005ge}.

The parameter space of MMAMSB models is given by
\bi
\item $m_{3/2}:\ 10-100$ TeV,
\item $\alpha:\ -15\rightarrow 15$,
\item $\tan\beta :3-60$,
\item $sign (\mu )=\pm$,
\ei
along with
\bi
\item $n_H,\ n_m=0,\ 1/2\ {\rm or}\ 1$.
\ei
Many of the following results can be understood by inspection of the $\alpha$ vs. $m_{3/2}$
plane plots available for each modular weight combination and shown in Ref. \cite{kklt3}.

\subsection{Cases with $n_H=0$}

\begin{figure}[tbp]
\includegraphics[height=0.28\textheight]{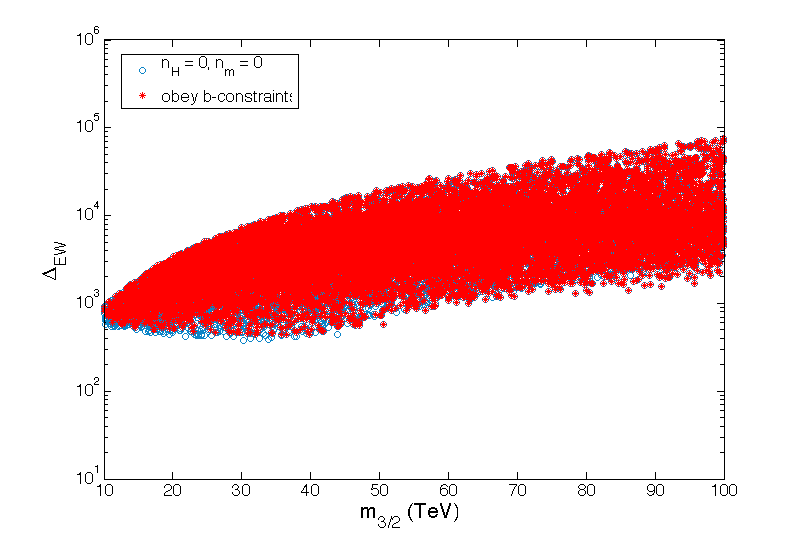}\\
\includegraphics[height=0.28\textheight]{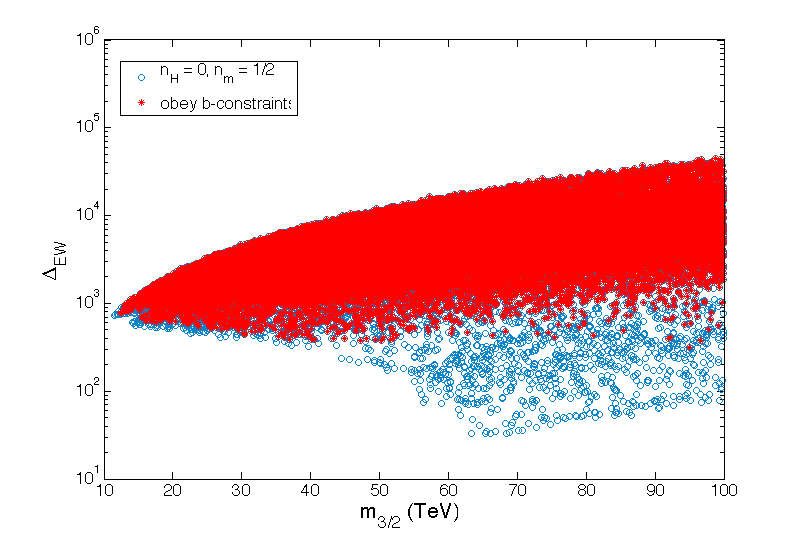}\\
\includegraphics[height=0.28\textheight]{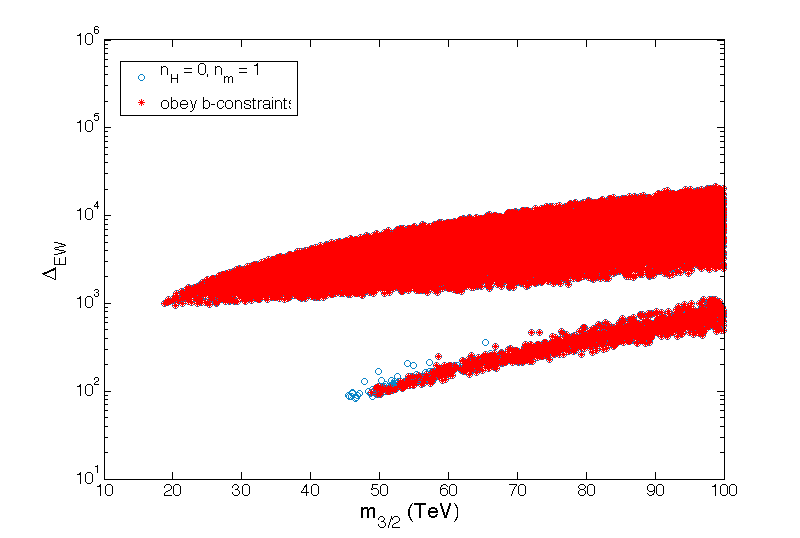}
\caption{Plot of $\Delta_{EW}$ vs. $m_{3/2}$ from a scan over MMAMSB parameter
space with $n_H=0$ whilst maintaining $m_h=125.5\pm 2.5$ GeV.
\label{fig:nH0}}
\end{figure}

Our first results for MMAMSB are shown in Fig. \ref{fig:nH0} in the $\Delta_{EW}$ vs. $m_{3/2}$ 
frame for cases with {\it a}). $(n_H,n_m)=(0,0)$, {\it b}). $(n_H,n_m)=(0,{1\over 2})$
and {\it c}). $(n_H,n_m)=(0,1)$.
For the case with $(n_H,n_m)=(0,0)$, we find a minimal value of $\Delta_{EW}\simeq 437$ 
which occurs at $m_{3/2}\sim 35$ TeV. 
At this point, $m_{\tg}\sim m_{\tq}\sim 1.8$ TeV which might be expected to be ruled out by 
LHC8 searches. However, the compressed spectra with gaugino masses 
$M_1,M_2,M_3\sim 800,1000,1800$ GeV leads to softer visible energy than might be otherwise
expected. The value of $\mu \sim 1200$ GeV produces the large value of $\Delta_{EW}$.
While much less tuned spectra are possible, they only occur with
very low $m_h$ values  and so are ruled out by the LHC8 Higgs discovery. 

The case with $(n_H,n_m)=(0,{1\over 2})$ has a minimal value of $\Delta_{EW}=314$ and so also is
EW fine-tuned. In this case, the minimum occurs for $m_{3/2}=95$ TeV which leads to $m_{\tg}=3.6$ TeV.
The large $\mu =1.1$ TeV value leads to high EW fine-tuning. 
Here, the LSP is the lightest Higgsino with mass $m_{\tz_1}\sim \mu$.
Models exist with $m_h\sim 125$ GeV and much lower fine-tuning reaching to $\Delta_{EW}\sim 30$
(blue points) but these all violate $B$-decay constraints due to rather light top squarks.

For the case with $(n_H,n_m)=(0,1)$, then the lowest $\Delta_{EW}$ value is found to be $\sim 91$, 
a considerable improvement but still nine times greater than the min from NUHM2. 
In this case, the solutions form two distinct branches-- the upper with $\alpha <0$
while the lower has $\alpha >0$. The lowest $\Delta_{EW}=91$ point actually has
$\mu\sim 150$ GeV, but with $m_{3/2}\sim 50$ TeV, then the top squarks have
mass $m_{\tst_{1,2}}\sim 2.1,\ 2.8$ TeV and not enough mixing so the values
of $\Sigma_u^u(\tst_{1,2})$ dominate the fine-tuning. 

\subsection{Cases with $n_H=1/2$}

\begin{figure}[tbp]
\includegraphics[height=0.28\textheight]{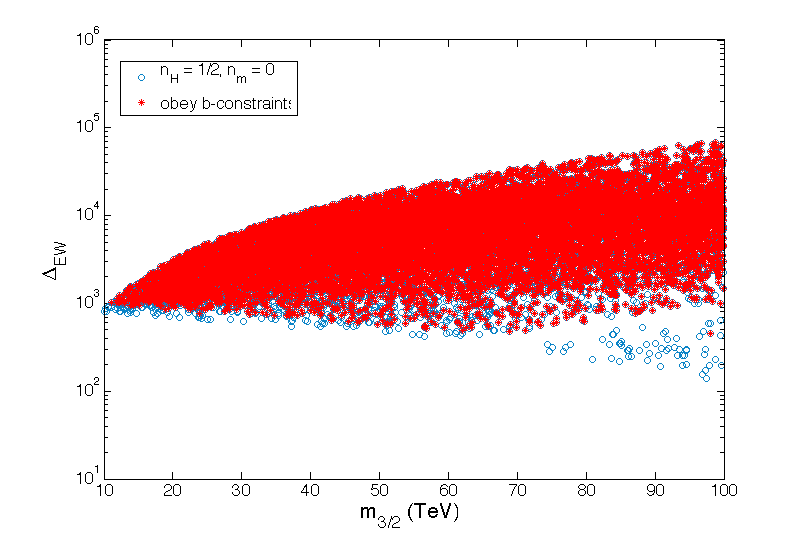}\\
\includegraphics[height=0.28\textheight]{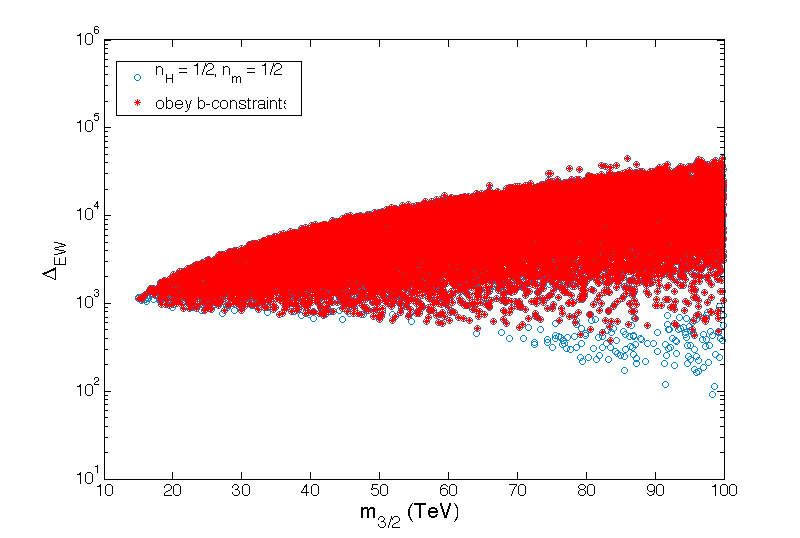}\\
\includegraphics[height=0.28\textheight]{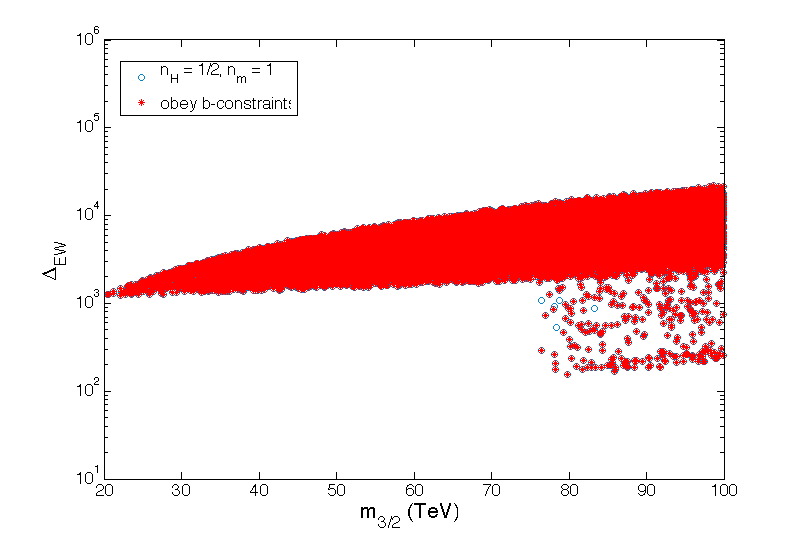}
\caption{Plot of $\Delta_{EW}$ vs. $m_{3/2}$ from a scan over MMAMSB parameter
space with $n_H=1/2$ whilst maintaining $m_h=125.5\pm 2.5$ GeV.
\label{fig:nH5}}
\end{figure}

Results for MMAMSB for cases with {\it a}). $(n_H,n_m)=({1\over 2},0)$, 
{\it b}). $(n_H,n_m)=({1\over 2},{1\over 2})$ and {\it c}). $(n_H,n_m)=({1\over 2},1)$
are shown in Fig. \ref{fig:nH5} in the $\Delta_{EW}$ vs. $m_{3/2}$ plane.
For frame {\it a})., we find a minimum $\Delta_{EW}=457$ at $m_{3/2}=98$ TeV
where a spectrum with $m_{\tg}\sim m_{\tq}\sim 2$ TeV but with $\mu =1.4$ TeV. 
The LSP is a neutral Higgsino with mass $\sim 1.34$ TeV and $\Omega_{\tz_1}h^2\sim 0.15$.
Even lower $\Delta_{EW}$ solutions reaching values of $\sim 100$ occur at very high
$m_{3/2}$, but these blue points are excluded by $B$-decay constraints.

For the $(n_H,n_m)=({1\over 2},{1\over 2})$ case in frame {\it b})., the lowest 
value of $\Delta_{EW}$ is found to be $375$ at $m_{3/2}=83$ TeV. 
Here again, $\mu\simeq 1.25$ TeV which gives a Higgsino-like LSP and rather compressed spectra.

For the $(n_H,n_m)=({1\over 2},1)$ case shown in frame {\it c})., then
$\Delta_{EW}$ can reach as low as $\sim 100$ at the high $m_{3/2}\sim 80$ TeV point.
For this point, $\mu$ drops as low as $589$ GeV and the LSP is again Higgsino-like
with a thermal underabundance of neutralino dark matter. The gluino and squark masses 
cluster around $3.5-4.5$ TeV, beyond LHC reach.

\subsection{Cases with $n_H=1$}

\begin{figure}[tbp]
\includegraphics[height=0.28\textheight]{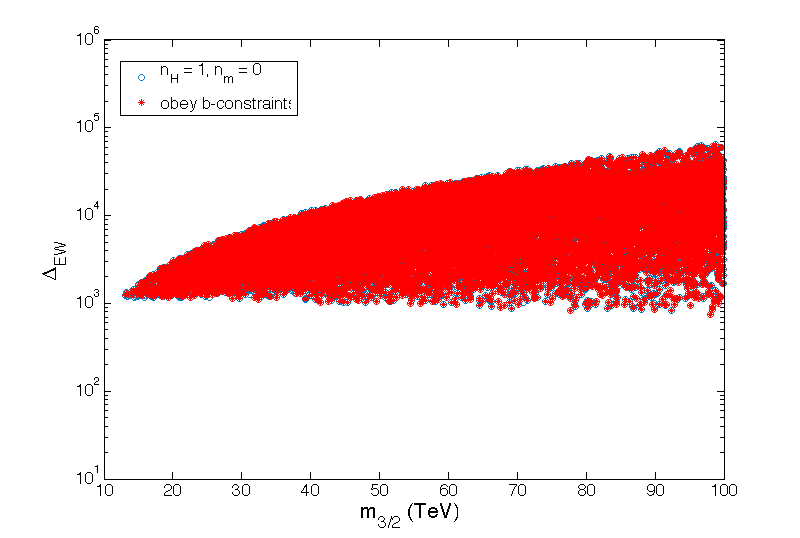}\\
\includegraphics[height=0.28\textheight]{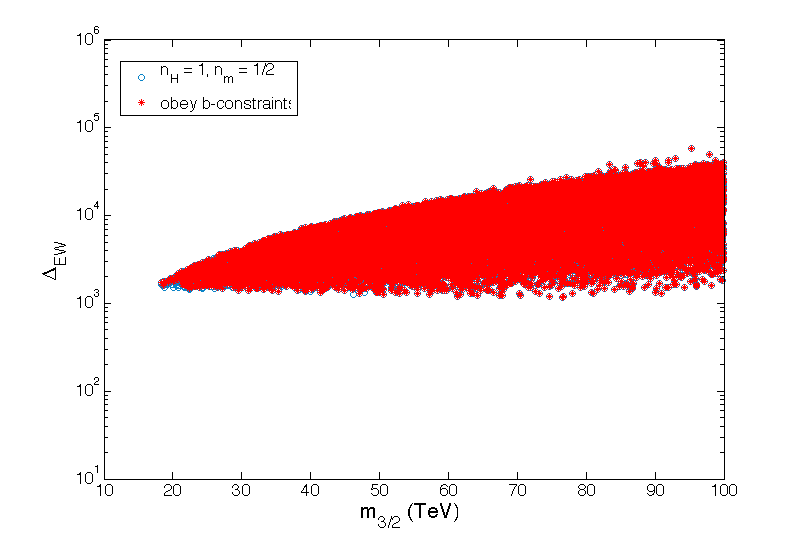}\\
\includegraphics[height=0.28\textheight]{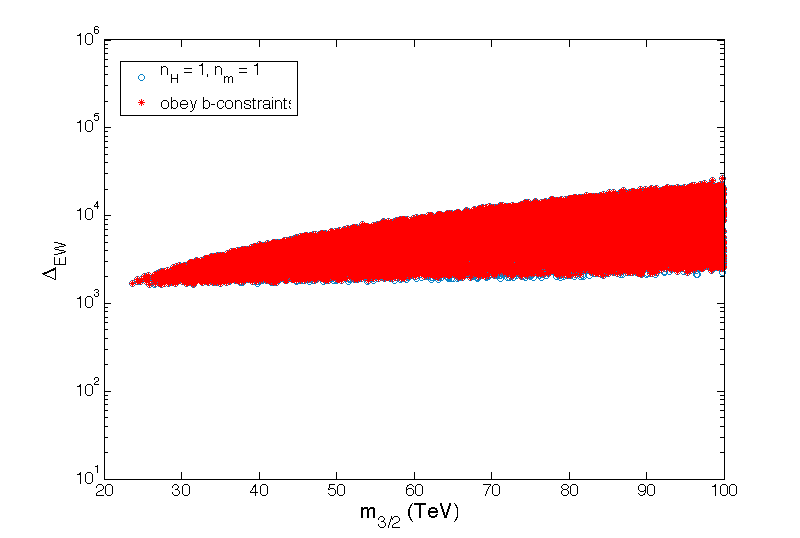}
\caption{Plot of $\Delta_{EW}$ vs. $m_{3/2}$ from a scan over MMAMSB parameters
space with $n_H=1$ whilst maintaining $m_h=125.5\pm 2.5$ GeV.
\label{fig:nH1}}
\end{figure}

The MMAMSB cases with {\it a}). $(n_H,n_m)=(1,0)$, {\it b}). $(n_H,n_m)=(1,{1\over 2})$ 
and {\it c}). $(n_H,n_m)=(1,1)$ are shown in Fig. \ref{fig:nH1}.
For the first case with $(1,0)$, then the min of $\Delta_{EW}$ is 859 at $m_{3/2}=90$ TeV.
The large EWFT is generated by the large $\mu =1.9$ TeV value.
Even so, the LSP is mainly bino with mass $m_{\tz_1}\sim 1.7$ TeV.

For frame {\it b}). with $(n_H,n_m)=(1,{1\over 2})$, then the min of $\Delta_{EW}$ is
1178 at $m_{3/2}=76$ TeV where $m_{\tg}\sim 3.9$ TeV and the bino-like LSP 
has mass $\sim 1.8$ TeV.

Finally, frame {\it c}). shows the $(n_H,n_m)=(1,1)$ case where a min of $\Delta_{EW}$
is found to be 1643 at $m_{3/2}=27$ TeV. Here, gluino and squark masses tend to exceed
$4$ TeV whilst $\mu\sim 2.6$ TeV which leads to the large EW fine-tuning.

\section{Conclusions:} 

In this paper, we have re-examined electroweak fine-tuning in light of 
recent LHC results on the Higgs discovery with $m_h\simeq 125.5$ GeV and the lack of 
any sort of signal for sparticles. This situation has lead to various claims that
the MSSM is no longer viable, or at least highly fine-tuned in the EW sector.
Alternatively, it has been claimed that conventional measures, applied conventionally,  overestimate EWFT.

To clarify the situation, we have proposed a Rule of Fine-tuning: 
When evaluating fine-tuning, it is not permissible to claim fine-tuning of 
{\it dependent} quantities one against another.
In the case of Higgs mass fine-tuning, we find that the measure $\Delta_{HS}$ violates this rule
by measuring non-independent terms which can lead to large cancellations. Then, 
Higgs mass fine-tuning can grossly overestimate-- often by orders of magnitude-- the electroweak
fine-tuning. By appropriately combining dependent terms, then $\Delta_{HS}$ reduces
to the model independent $\Delta_{EW}$ measure: the offending large logs are still present, but
can now cancel against other non-independent terms.

We have also examined the traditional measure $\Delta_{BG}$.
In this case, the measure appears at first sight to be highly model-dependent. 
The model-dependence is traced to the fact that most users regard the 
multiple parameters of most popular SUSY models as {\it independent} 
when in fact their independence is only an artifact of trying to
construct a model which encompasses a wide range of hidden sector 
SUSY breaking possibilities. If instead one relates the various soft parameters-- such as
multiples of $m_{3/2}$ as expected in supergravity models with SUSY broken via the 
superHiggs mechanism-- then it is shown that $\Delta_{BG}$ also reduces to the
model-independent electroweak measure $\Delta_{EW}$. 

For low $\Delta_{EW}$, then it is required that 1. $\mu\sim 100-300$ GeV, 2. $m_{H_u}^2$ is radiatively driven to
small negative values $\sim m_Z$ and 3. the top-squarks are in the few TeV range with
large mixing. The large mixing reduces top-squark radiative contributions to $\Delta_{EW}$
while lifting $m_h$ into the 125 GeV range.

\begin{figure}[tbp]
\includegraphics[height=0.3\textheight]{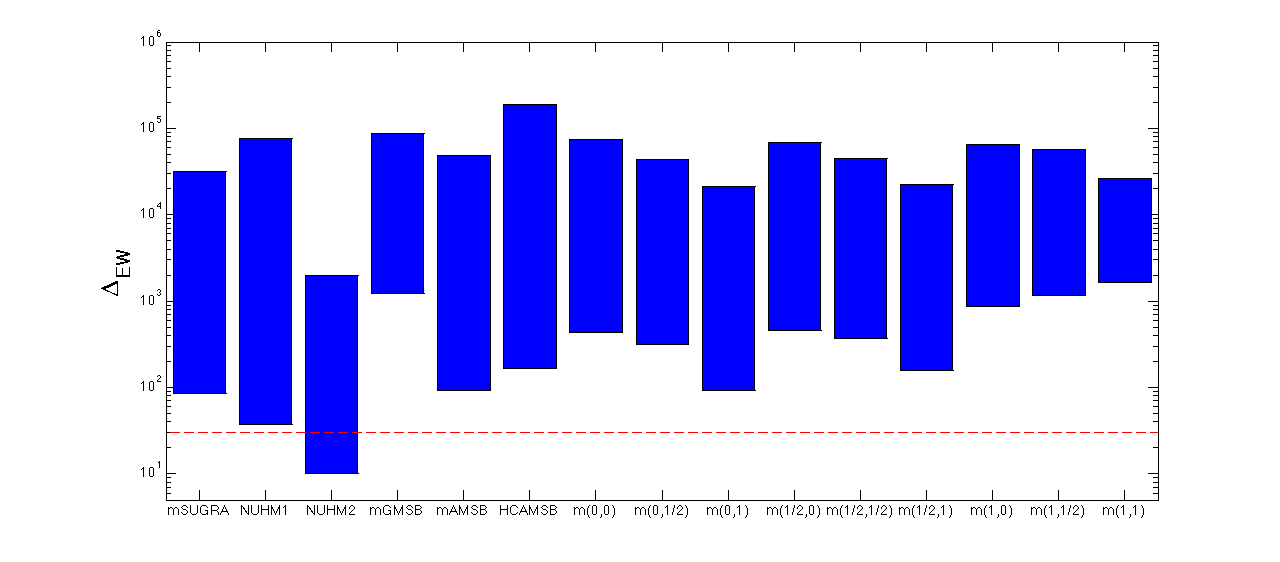}
\caption{Histogram of range of $\Delta_{EW}$ values generated for each SUSY model
considered in the text. We would consider $\Delta_{EW}\alt 30$-- the lower the better--
as acceptable values for EW fine-tuning. This region is located below the dashed red line.
\label{fig:histo}}
\end{figure}

We also evaluated $\Delta_{EW}$ values from a scan over parameters of 15 models:
mSUGRA, NUHM1, NUHM2, mGMSB, mAMSB, HCAMSB and nine cases of mixed moduli-anomaly
(mirage) mediated SUSY breaking. Our overall results are summarized in 
Fig. \ref{fig:histo} where we show the range of $\Delta_{EW}$ generated on the $y$-axis
versus models on the $x$-axis. Only one model-- NUHM2-- reaches to the rather low
$\Delta_{EW}\sim 10$ values, indicating just 10\% EWFT. This can be so because
the freedom in the soft Higgs sector allows arbitrarily low values of $\mu$ 
(subjectto LEP2 constraints) to be 
generated while at the same time driving $m_{H_u}^2$ to just small negative values,
while also accommodating TeV-scale top squarks with large mixing. For the remaining
models, their inherent constraints make satisfying these conditions with 
$m_h\sim 125$ GeV very difficult unless they are highly fine-tuned. 
The best of the remainder models include $NUHM1$ which allows for min $\Delta_{EW}$
as low as 30. Thus, $\Delta_{EW}$ does indeed put SUSY models under seige. 

Luckily, at least NUHM2 and its generalizations survive, and even thrive. 
In the case of the surviving NUHM2 spectra (those with $\Delta_{EW}\alt 30$),
a discovery at LHC14 might take place provided $m_{\tg}\alt 2$ TeV\cite{lhc}: 
this reach covers about half of parameter space\cite{rns}. The definitive search
for SUSY would have to take place at a linear $e^+e^-$ collider where
$\sqrt{s}$ could extend beyond $2m(higgsino)$-- in this case $\sqrt{s}\sim 500-600$ GeV
is required for a thorough search.\footnote{The proposed TLEP $e^+e^-$ collider
with projected maximal $\sqrt{s}\sim 350$ GeV may not have sufficient energy to 
thoroughly explore natural SUSY\cite{tlep}.}
Such a machine would either discover SUSY or rule out SUSY naturalness\cite{snowmass2}.
We may also expect an ultimate discovery of a Higgsino-like WIMP and 
a DFSZ-type axion, since models such as SUSY DFSZ solve the strong CP fine-tuning problem
and the $\mu$ problem while at the same time allowing naturally for a Little Hierarchy of
$f_a\ll m_s$, where $m_s\sim 10^{11}$ GeV represents the mass scale 
usually associated with hidden sector SUSY breaking. 
That hierarchy is then reflected by the hierarchy $\mu\ll m_{3/2}$
which seems to be what naturalness combined with LHC data is telling us.

We end by confessing that the general features of some of our viewpoints have been 
articulated previously by Giudice\cite{gian}:
\begin{quotation}
``It may well be that, in some cases, Eq. \ref{eq:DBG} overestimates the amount of tuning. 
Indeed, Eq. \ref{eq:DBG} measures the sensitivity of the prediction of $m_Z$ 
as we vary parameters in ``theory space''. 
However, we have no idea how this ``theory space'' looks like, and the procedure of
independently varying all parameters may be too simple-minded.''
\end{quotation}
Amen!

\section*{Acknowledgments}

We thank P. Huang, D. Mickelson, A. Mustafayev and X. Tata for many 
previous collaborations leading up to this study.
HB would like to thank the DESY Helmholtz Alliance group for hospitality while this work was completed.
This work was supported in part by the US Department of Energy, Office of High
Energy Physics.

%

%
\end{document}